\documentstyle[aasms4,psfig]{article}

\newcommand\beq{\begin{equation}} 
\newcommand\eeq{\end{equation}}

\received{} 
\accepted{} 

\lefthead{Wu et al.} 
\righthead{Surface Photometry of the Edge-on Galaxy: NGC 4565} 

\begin{document}

{\title{Intermediate-band Surface Photometry of the Edge-on Galaxy: NGC 4565}

\author{ 
Hong Wu\altaffilmark{1}, 
David Burstein\altaffilmark{2},
Zugan Deng\altaffilmark{1,3},
Xu Zhou\altaffilmark{1},
Zhaohui Shang\altaffilmark{4},
Zhongyuan Zheng\altaffilmark{1},
Jiansheng Chen\altaffilmark{1},
Hongjun Su\altaffilmark{1},
Rogier A. Windhorst\altaffilmark{2},
Wen-ping Chen\altaffilmark{5},
Zhenlong Zou\altaffilmark{1},
Xiaoyang Xia\altaffilmark{1,6},
Zhaoji Jiang\altaffilmark{1},
Jun Ma\altaffilmark{1},
Suijian Xue\altaffilmark{1},
Jin Zhu\altaffilmark{1},
Fuzhen Cheng\altaffilmark{7,1} ,
Yong-Ik Byun\altaffilmark{5,8},
Rui Chen\altaffilmark{1},
Licai Deng\altaffilmark{1},
Xiaohui Fan\altaffilmark{9},
Li-Zhi Fang\altaffilmark{10},
Xu Kong\altaffilmark{7,1},
Yong Li\altaffilmark{2},
Weipeng Lin\altaffilmark{1,11},
Phillip Lu\altaffilmark{12},
Wei-hsin Sun\altaffilmark{5},
Wean-shun Tsay\altaffilmark{5},
Wen Xu\altaffilmark{2},
Haojing Yan\altaffilmark{2},
and Zheng Zheng\altaffilmark{13}
}

\altaffiltext{1}{National Astronomical Observatories, 
    Chinese Academy of Sciences, Beijing 100012, China }

\altaffiltext{2}{Department of Physics and Astronomy, Box 871504, Arizona
    State University, Tempe, AZ  85287--1504}

\altaffiltext{3}{Graduate School, Chinese Academy of Sciences, 
    Beijing 100080, China}

\altaffiltext{4}{Department of Astronomy, University of Texas at Austin,
    Austin, TX 78712}

\altaffiltext{5}{Institute of Astronomy, National Central University,
    Chung-Li, Taiwan, China}

\altaffiltext{6}{Department of Physics, Tianjin Normal University, China}

\altaffiltext{7}{Center for Astrophysics, University of Science and
       Technology of China, Hefei, 230026, China}

\altaffiltext{8}{Center for Space Astrophysics and Department of
Astronomy, Yonsei University, Seoul, 120--749, Korea}

\altaffiltext{9}{Princeton University Observatory, Princeton, New
    Jersey, 08544}

\altaffiltext{10}{Department of Physics, University of Arizona, Tucson,
    AZ 85721}

\altaffiltext{11}{ The Partner Group of MPI f\"ur Astrophysik,
    Shanghai Astronomical Observatory, Shanghai 200030, China}

\altaffiltext{12}{Department of Physics and Astronomy, Western 
    Connecticut State University, Danbury, CT 06810}

\altaffiltext{13}{ Department of  Astronomy, Ohio State University,
    Columbus, OH 43210}

\authoremail{wu@vega.bac.pku.edu.cn}

\begin{abstract}

We present a deep, 42.79 hr image of the nearby, edge-on galaxy NGC
4565 in the Beijing-Arizona-Taipei-Connecticut (BATC) $\rm 6660\AA$
band using the large-format CCD system on the 0.6m Schmidt telescope
at the Xinglong Station of the National Astronomical Observatories of
China (NAOC).  Following the procedures previously developed by our
team for the analysis of deep images of galaxies (Zheng et al.), we
obtain a final image that is calibrated to an accuracy of 0.02 mag in
zero point, and for which we can measure galaxy surface brightness to
an accuracy of 0.25 mag at a surface brightness at 27.5 mag
arcsec$^{-2}$ at 6660$\rm \AA$, corresponding to a distance of 22 kpc
from the center of the disk.  The integrated magnitude of NGC~4565 in
our filter is $m_{6660} = 8.99$ (= R magnitude of 9.1) to a surface
brightness of 28 mag arcsec$^{-2}$. We analyze the faint outer parts
of this galaxy using a two-dimensional model comprised of three
components: an exponential thin disk, an exponential thick disk, and a
power-law halo. Combined with a need to provide a cut-off radius for
the disk, a total of 12 parameters are included in our model.  We
determine the best values of our model parameters via 10,000 random
initial values, 3,700 of which converge to final values.  We then plot
the $\chi^2$ for each converged fit versus parameter value for each of
the 12 parameters.  The thin disk and thick disk parameters we
determine here are consistent with those of previous studies of this
galaxy.  However, our very deep image permits a better determination
of the power law fit to the halo, constraining this power law to be
between $r^{-3.2}$ and $r^{-4.0}$, with a best fit value of
$r^{-3.88}$.  We find the axis ratio of the halo to be 0.44 and its
core radius to be 14.4 kpc (for an adopted distance of 14.5 Mpc).  We
also agree with others that the bulge of NGC~4565 is fit well by an
exponential luminosity distribution with scale height similar to that
found for the thin disk.
\end{abstract}


\section{INTRODUCTION}

In this paper we continue our investigations into the faint surface
brightness distributions of edge-on galaxies using data obtained 
as part of the Beijing-Arizona-Taipei-Connecticut (BATC) Multi-Color 
Survey of Sky (cf. Fan et al. 1996, Yan et al. 2000). In previous
papers detailing our investigations of NGC~5907 (Shang et al. 1998; 
Zheng et al. 1999), we showed that this galaxy does not have a 
luminous halo, counter to what was previously suggested (Sackett et
al. 1994; Morrison, Boroson, \& Harding 1994).  Rather, our images
showed a faint ring around NGC~5907, likely the result of the tidal
disruption of a dwarf galaxy.

The existence of very faint surface brightness features around edge-on
spiral galaxies is further investigated here with BATC observations of
the well-known galaxy NGC~4565.  As opposed to NGC~5907, NGC~4565 is
at high Galactic latitude ($86.44^\circ$), implying that its halo
should be less contaminated by bright Galactic stars than are our
observations of NGC 5907. NGC~4565 is classified as Sb (\cite{RC3}). 
We place it at a distance of 14.5 Mpc, based on its distance in the
Mark III catalog (1043 km s$^{-1}$; \cite{Will96}) and a Hubble
constant of 72 km s$^{-1}$ $\rm Mpc^{-1}$ (\cite{Freed01}).  It is
known to have a Seyfert nucleus (\cite{Ho97}), and has been much
studied in the past in terms of optical surface photometry
(\cite{JT82}, hereafter JT; \cite{K79}; \cite{KS81}, hereafter KS;
\cite{KB78}; \cite{H80}; \cite{NJ97}, hereafter NJ; \cite{DW86}).
Most of those earlier studies were based on photographic data.  The
V-band data of NJ used a CCD with a relatively small field of view,
making it difficult for them to accurately determine sky background
levels.  Infrared J, H and K imaging by Rice (1996) completes the
existing photometric imaging data on this galaxy.

Our observations and the details of our reduction of the data we have
obtained for NGC~4565 are given in Section 2.  The measurement of
the luminosity profiles and error analysis are presented in Section
3. Section 4 gives the results of model fitting, comparison and
possible systematic effects from PSF and disk inclination.  The 
last section summarizes the main results of this paper.

\section{OBSERVATIONS AND DATA REDUCTION}

\subsection{Observations}

Observations of NGC 4565 were obtained with the 60/90cm Schmidt
telescope at the Xinglong Station of the National Astronomy
Observatories of China (NOAC), using a thick Ford $\rm 2048\times2048$
CCD with 15 {$\mu$}m pixels at the f/3 prime focus.  The field of view
of this CCD is $58' \times 58'$ and the scale is 1.7$''$/pixel.  With
the nearly one degree field of view, there is sufficient sky in a
single frame such that objects with visual sizes less than 30$'$ can
have their sky background determined accurately.  The Lick data-taking
system is employed and all the CCD images are overscan-subtracted
(i.e., initially bias-subtracted) during the readout time 
(\cite{Zheng99}).

The filter used for the observations reported here is the BATC filter
with central wavelength of 6660{$\rm\AA$} and bandwidth of $\rm
480{\AA}$ (\cite{Fan96} and \cite{Yan00}). This filter is a good
compromise between getting as far to the red as possible, and avoiding
many bright sky emission lines with the broadest possible
filter. Indeed, all of the BATC filters are designed to avoid
contamination by emission lines from the night-sky (cf. Fan et
al. 1996).

We obtained 190 images of NGC~4565 from 1995 to 1997, of which
we deemed 150 images taken in 22 runs (Table 1) as suitable
for analysis.  Most of these images are of exposure times 900 to
1200 sec.  In a given night all exposures were dithered randomly
at a level of $\sim 10$ pixels to facilitate removal of cosmic
rays and CCD defects during the data reduction.  The net effect
of such dithering is to reduce the full exposure field of view
to less than the full field of the CCD.  All images used in this 
analysis have FWHM seeing between 1.7 to 2.6 pixels, or 2.9$''$ 
to 4.4$''$, somewhat larger than the typical seeing at the
Xinglong observing station.  During the observations the gain of the
CCD system was adjusted several times (for various reasons owing to
having the CCD system work well), resulting in gains of 3.7
e$^{-}$/adu in 1995, 4.1 e$^{-}$/adu in 1996, and 3.3 e$^{-}$/adu in
1997.  Readout noise was constant (12 e$^{-}$) during the three years.

\begin{table}[]
\caption[]{Brief observation log of NGC 4565}
\vspace {0.5cm}
\begin{tabular}{lrrr}
\hline
\hline
Date   &   N\tablenotemark{a}  & Exp.(sec.)\tablenotemark{b}  
&  Gain(e$^-$/adu) \\
\hline
95/01/28   &     1     &     1800      &         3.7 \\
95/03/03   &     8     &     9600      &         3.7 \\
95/03/04   &     2     &     1800      &         3.7 \\
95/05/05   &     2     &     2400      &         3.7 \\
95/05/06   &     7     &     6623      &         3.7 \\
95/05/24   &     4     &     3720      &         3.7 \\
96/01/21   &     4     &     4800      &         4.1 \\
96/02/17   &    17     &    15300      &         4.1 \\
96/02/18   &    20     &    18000      &         4.1 \\
96/02/19   &    20     &    18000      &         4.1 \\
96/02/20   &    16     &    14400      &         4.1 \\
96/02/21   &     4     &     3600      &         4.1 \\
97/03/15   &     4     &     4800      &         3.3 \\
97/03/29   &     3     &     3600      &         3.3 \\
97/03/30   &     1     &     1200      &         3.3 \\
97/04/04   &     4     &     4800      &         3.3 \\
97/04/05   &     6     &     7200      &         3.3 \\
97/04/08   &     5     &     6000      &         3.3 \\
97/04/09   &     7     &     8400      &         3.3 \\
97/04/10   &     5     &     6000      &         3.3 \\
97/04/11   &     6     &     7200      &         3.3 \\
97/05/28   &     4     &     4800      &         3.3 \\
           &           &               &         \\  
Total      &   150     &   154043      &         \\
\hline
\end{tabular}\\
$^a$ -Numbers of frames observed \\
$^b$ -The total exposure time at one night \\
\end{table}

\subsection{Bias and dark}

The mean value of overscan-subtracted bias was stable within any one
month period of time. 10 bias frames (5 at the start of the night, 5
at the end) are taken daily for the BATC program.  As we can track the
stability of the bias during the year, we are able to average from 200
to 300 individual bias frames for each night of observation, which
removes any stable structure in the bias frame (any variable level is
removed via the overscan subtraction).  It is this average of many
bias frames that is subtracted from the images of a given night. 
The CCD dark count has been constantly monitored throughout the BATC 
survey, and has always been found to be stable and rather free of 
gradients.  As the average dark count/pixel is 3 e$^-$/hour, the 
dark count level in a 42.8 hour exposure is 128 e$^-$/pixel, or 
0.04\% of the sky level in this combined image and the spatial 
variation of dark is even smaller; i.e., of negligible importance.
This constant dark count value was scaled to exposure time for 
each image, then subtracted.

\subsection{Flat Field}

As detailed in our previous papers (cf. Fan et al. 1996; Zheng et al.
1999), the BATC program has developed the means by which we can use
dome flats to obtain accurate, high signal-to-noise (S/N), flat fields
for flat-fielding sky images.  Briefly, the Xinglong Schmidt telescope 
is equipped with a UV-transparent plastic diffuser plate that can be 
firmly placed directly in front of the Schmidt corrector. The diffuser
provides randomly scattered light to the corrector, reproducing the
flux from a uniform sky.  Such a method is necessary to accurately flat
field our images, as Wild (1997) points out that over a one degree
scale, no part of the sky is really flat.  Additionally, this fact is
attested to by the many BATC images our survey has obtained.  We note,
however, that it is only with narrow or intermediate-band filters that
this diffuser on a Schmidt telescope can produce reliable flat fields.
Otherwise, the dome flat field can introduce second-order color terms 
to broad band observations that must be removed using direct sky
images (\cite{Fan96}).

We take twelve dome flats each day, each with exposure times of 150
seconds. This length of exposure reduces the effect of the finite time
for shutter opening and closing, resulting in a spurious gradient of
less than 0.013\% (cf. Sec. 3.3).  All twelve dome flats in one day
are combined as the final flat field to correct the frames obtained on
the same night. The total count of combined dome flat is about 840,000
electrons per pixel, far higher than that of sky background (about
2,000 electrons per pixel) in a single exposure frame.

\subsection{Image Combination}

Air mass corrections have to be done for our images on a
pixel-by-pixel basis, as there exists close to a 1\% gradient in air
mass correction over one square degree, even at an altitude of 60
degrees.  To take into account the different sets of data (gain
differences, dithering, seeing differences, etc.), we combine these
images in two steps.  First we separate the image frames into 11
groups. Each group includes the frames observed in the same state of
instrument system and similar observation conditions (e.g., similar
seeing).  On average, the observations for a given night are in the
same group.  The availability of hundreds of well-defined point
sources (both stars and distant galaxies) in these frames aided the
combination of all frames to a common system.  This combination
accounts for both dithered frames as well as for slight rotations in
the CCD chip from year-to-year.  To get all combined images to the
same effective seeing radius of 2.3 pixels ($4"$), some of the
combined images were convolved with Gaussian functions to add small
additional values of seeing.  Three sigma rejection was used to remove
cosmic-rays, hot-points, bad pixels and Schmidt telescope-related
ghost-images.  The final image shown in Figure~1 is the result of
merging the 11 combined frames, a total of 42.79 hours of observation.

\subsection{Flux calibration}

Photometric calibration was provided by the five nights that were
photometric (Mar. 4, 1995, Mar. 6, 1995, Jan. 6, 1997, Jan. 17, 1997,
and Apr. 16, 1997).  Following the now-standard BATC photometric
calibration (cf. \cite{Zhou01}; \cite{Yan00}), four Oke \& Gunn (1983)
standard stars (HD 19445, HD 84937, BD +26$^{\rm \circ}$2606 and
BD+17$^{\rm \circ}$4708) are used as BATC calibration stars.  The
calibrations of five nights agree quite well, yielding a zero point
accuracy of 0.02 mag as determined from 48 bright stars over the
five photometric nights. This zero point yields a sky background in 
the 6660$\rm \AA$ intermediate-band BATC filter of 20.72 mag $\rm
arcsec^{-2}$, with a corresponding magnitude of 20.30$\pm$0.02 mag 
for 1 electron $\rm second^{-1}$ in the 42.79 hour image.  The sky
background is 0.54 mag brighter than it was for our NGC~5907
observations with the same filter (cf. Zheng et al. 1999), likely
owing to a combination of most of the images being taken closer to
solar maximum (1996-1997) than were the NGC~5907 images, plus an
increase of sky brightness over the last few years at the Xinglong
observing station (\cite{Liu01}).

\section{MEASUREMENT OF PROFILES AND ERROR ESTIMATE}

Accurate determination of sky background level is a key point for
accurate, deep surface photometry.  Small variations in sky
background, either due to undulations in the CCD sensitivity, or star
halos, can affect the measurement of very faint surface brightnesses
in galaxies. These issues have been investigated in detail for our
data-taking system in our previous investigation of NGC~5907 (Zheng et
al.).  Here we employ the same methodology as Zheng et al. in
obtaining as accurate a fit to the sky background as our data allow.

We first trim our final image to a size of $\rm 1701\times1701$ pixels
($\rm 48.2'\times48.2'$) centered on NGC~4565 (Figure 1).  The small
size, relative to our full CCD, is due to the dithering of many
images.  The average sky level in the final image is close to 302,500
e$^-$/pixel, which would nominally yield a statistical error of 550
e$^-$/pixel.  However, in combining the images the data are effectively
smoothed in to an area of somewhat less than 4 pixel$^2$, yielding a
statistical error of 320 e$^-$/pixel for sky in the combined image.  In
neither case, however, does this statistical error reflect the true
error in sky background value, as the true error is convolved with
other large-scale sources of variation on the CCD (cf. disucssion below).  

\begin{figure}
\figurenum{1}
\caption[]{The full BATC 6660$\rm \AA$ final image of NGC~4565,
comprised of 42.79 hours of observation.  The field of view is 1701
$\times 1701$ pixels, or 48.2$' \times$48.2$'$ in size.
\label{fig1}}
\end{figure}

\subsection{Star and galaxy masking}

The software SExtractor (\cite{Bertin96}) was used to find all objects
in our final image with peak flux higher than 3 sigma above the sky
background, as well as to separate stars from some of the fainter
galaxies in this field.  Following Zheng et al., we also classified
all stellar--appearing objects into four types: point sources (stars
and faint galaxies) whose wings extend less than 25 pixels at a mask
level of 40 e$^{-}$ (about 10\% the sky variation), bright stars (wing
radii $>$ 25 pixels), saturated stars (whose peaks are higher than
10$^7$ e$^{-}$) and resolved galaxies.  We find more than 12,000
objects in the trimmed image, similar to what we found for the
NGC~5907 long exposure images.

To remove as much of the wings of the bright stars as possible, we
construct an average PSF from the PSFs of nearly 100 relatively
isolated bright stars.  In so doing, we can reduce the size needed to
mask bright stars by first removing the fainter parts of their PSFs,
which are fit using the average PSF.  Unfortunately, at the very faint
light levels to which our data are sensitive ($\sim 29$ mag
arcsec$^{-2}$), we find nearly 10,000 faint objects (almost all
distant galaxies at this high Galactic latitude; cf. Odewahn et al. 1996), 
which limits the radius of our accurate PSF to 25 pixels.

The PSFs of the myriad of faint objects are well fit by this average
PSF within a radius of 25 pixels.  The fact that we have to restrict
the PSF to a radius of 25 pixels is unfortunate for, as luck will have
it, too many of the bright stars in this image seem to be near this
galaxy (cf. Fig. 1).  As shown in Zheng et al., residual light from
bright stars affects the measurement of very faint light levels around
galaxies.  On the other hand, choosing very large masks is not an option,
as we find that if we masked out the full PSF of the saturated stars
near NGC~4565, we would be left with little of the galaxy to measure.

As a compromise, we first assume that the outer wings of bright and
saturated stars are center-symmetric, then later deal with the known
asymmetries in the Schmidt PSF.  We apply median filters to construct
a one-dimensional PSF, whose wings can extend to as far as 300 pixels
(510 arcsec) in radius.  From this symmetric PSF we construct a
two-dimensional PSF and separately subtract it from all the bright and
saturated stars.  We subtract the wings of the PSF from saturated
stars manually to find the best scale factor for each star.  The final
masks for the saturated stars are designed to be larger than the
residual asymmetrical ripples seen in the subtracted PSFs
(cf. \cite{Zheng99}). Also sets of small masks are applied to mask
the spikes in the Schmidt PSF.  Background galaxies and faint stars
are masked entirely, with the radius of the masks checked by eye.
NGC~4565 itself was masked using three circles: one with a radius of
350 pixels, and two with radii of 250 pixels, The final masked image
is shown as Figure~2.

\begin{figure}
\figurenum{2}
\caption[]{This is the full mask applied to the 6660$\rm \AA$ 
final image of NGC~4565, blocking out all point sources as well as
a large region about the galaxy itself.
\label{fig2}}
\end{figure}

\subsection{Sky background fitting}

The remaining sky background comprises 27.3\% of the area of the final
masked image.  Before sky fitting, a median filter of 15$\times$15
pixels is convolved with the masked image using only unmasked pixels.
This was done to increase the fraction of area with sky background, as
most of masked regions with radii less than 10 pixels are replaced by
surrounding background.  After such median filtering, the effective
sky background increases to 52.8\% of the whole image.
 
We use the method of Zheng et al. to fit the sky background. Briefly,
because of the large masked central region of object galaxies,
one-dimensional row or column fitting with functions of high order
will introduce uncontrolled fluctuations in the fitted sky.  To 
avoid this, a one order Spline3 fit was applied in one-dimension, 
both row-by-row and column-by-column separately (cf. Zheng et al.), 
using only those unmasked points with backgrounds $<2.5 \sigma$ of 
the average.  We then average the row-fitted and column-fitted 
images and smoothed the average with a circular-gaussian filter 
of $\sigma$=30 pixels. This smoothed fit is then adopted as the 
sky background and subtracted from the final image.

As a test, we also use one-dimensional, low-order Legendre polynomials
in place of the Spline3 fits.  The difference between the two sky 
images in the central region around NGC~4565 is less than 40 $\rm
e^{-} pixel^{-1}$ (out of a sky background of 302,500 $\rm e^{-}
pixel^{-1}$). Hence, determination of the sky background is not
significantly affected by the spline-fitting function used. To 
estimate the systematic error of sky-fitting, we again mask the 
background subtracted image and separate the image into 400 
adjoining regions of 85$\times$85 pixels. The standard variation 
of mean values of these 85$\times$85 boxes is 115 $\rm e^{-} 
pixel^{-1}$.  This is regarded as the best estimate of the added 
error introduced by our sky-fitting procedure.  Figure~3 shows 
the plots of four slices of sky background--subtracted image, 
which shows the accuracy of the method (cf. Figures 7 and 8 in 
Zheng et al.).

\begin{figure}
\figurenum{3}
\centerline{\psfig{file=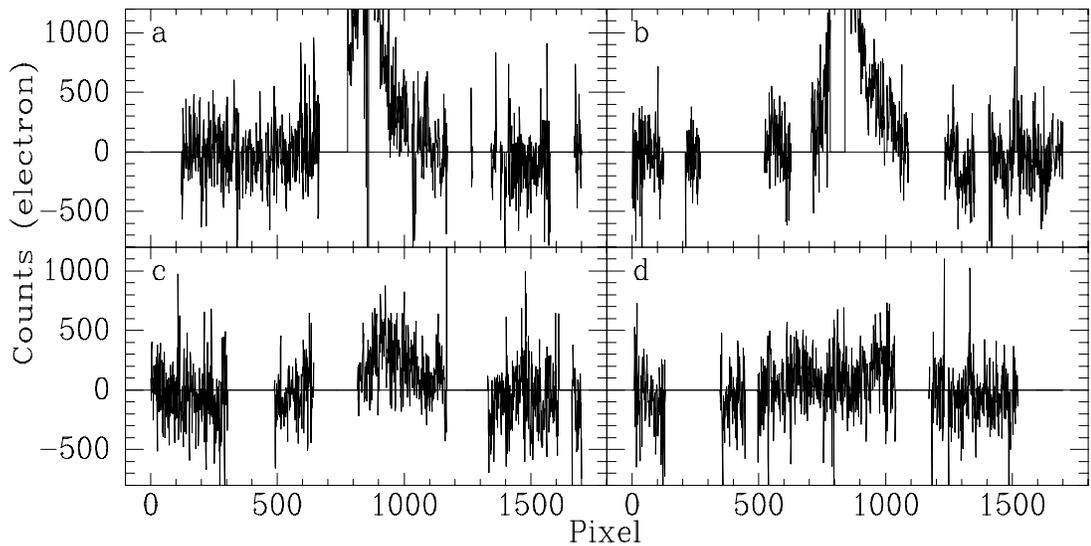,width=16.0cm}}
\caption[]{Four slices of the background-subtracted 6660$\rm\AA$ image
to show the accuracy of the sky background fit. Each slice is averaged
over 35 lines of the image, and only non-masked pixels are used. If
all 35 points are masked, a zero value is assigned. The straight lines
indicate the zero level. a) 8 arc-minute SE of the minor axis. b) 8
arc-minute NW of the minor axis. c) 8 arc-minute SW of the major
axis. d) 8 arc-minute NE of the major axis. The large scale variation
of background subtraction is small.  The disk of NGC~4565 is seen
in the centers of slices (a) and (b). The wings of subtracted stars
are also evident in these slices.
\label{fig3}}
\end{figure}

\subsection{Error estimate}

There are two types of errors that can affect our results. One type is
random and yields a Poisson distribution that can be suppressed by
increasing the number of sampled points.  The another type is
systematic in nature, and is usually independent of sample size. In the
following, we will discuss the error sources individually.

{\it Readout noise}

Readout noise is random noise which is introduced in the process of
data readout.  For our CCD, the value is 12 $\rm e^{-} pixel^{-1}$ per
frame.  Because the combined image is made of 150 frames, the final
noise is 147.0 $\rm e^{-} pixel^{-1}$.  The bins used for the analysis
of the light distribution in the halo of NGC~4565 vary in size from 24
pixels (a box of 8$\times$3 pixels) to 1750 pixels (a box of 35
$\times$ 50 pixels).  Hence, when averaged in this way, the readout
noise is 30.0 $\rm e^-$ for the smallest bins, and 3.5 $\rm e^-$ for
the largest bins.

{\it Sky background photon noise} 

As discussed above, the merging process to produce the final
image effectively smooths the data, such that the flux in each pixel
is an average of nearly 4 pixel$^2$ around it.  Hence, in the final
image, the photon error per pixel is 320 e$^{-}$, as opposed to the
550 e$^{-}$ one might otherwise expect from a mean sky background of
302,500 $\rm e^{-} pixel^{-1}$.  Using this value, we estimate the
photon sky noise to be 15 e$^{-}$ for a region $\rm 35 \times
50 $ pixels in size (our largest bin size), and 130.6 e$^-$ for a
region $\rm 8\times3$ pixels in size (our smallest bin size).
  
{\it Dark current noise} 

The dark current of our combined image of 42.79 hour is 128.3 $\rm e^-
pixel^{-1}$.  This introduces random noise of 11.3 $\rm e^{-}
pixel^{-1}$. For the smallest and largest bins, this value reduces to
2.3 $\rm e^- $ and 0.27 $\rm e^- $ separately.

{\it Bias subtraction noise}

Since the readout noise is 12 $\rm e^{-} pixel^{-1}$ and 200 to 300
bias frames are used to form final average bias, the error from bias
subtraction is about 0.8 $\rm e^{-} pixel^{-1}$, yielding 0.02 $\rm
e^{-} pixel^{-1}$ for the region of $\rm 35 \times 50$ pixels and 0.17
$\rm e^-$ for the smallest bins used here.

{\it Flat field: Random and Systematic Noise}

The averaged nightly flat used here is comprised of 12 dome-flats with
total counts about 840,000 $\rm e^-/pixel$, yielding a photon noise
close to 0.1\%, or 330.1 $\rm e^{-} pixel^{-1}$ equivalent sky background
counts.  Given that that final image includes 150 shifted, averaged
flats the error per pixel is reduced to 0.009\%. This part of the
flat field error is random in nature and translates to a formal error of
0.0002\% for the largest sample bins used here, to 0.0018\% for the 
smallest bins.

The stability of the large-scale flat field was checked by comparing
the dome-flats from adjacent nights. The average results of these
tests indicate an error of 0.03\%, or close to 90 $\rm e^{-}
pixel^{-1}$ for our measured sky background.  In our CCD system, the
opening and closing time of shutter is 20 milliseconds and the
exposure time of each dome-flat is 150 seconds. So there exists a
0.013\% gradient due to finite shutter speed in the each dome-flat
frame. Combining these two sets of errors, we estimate the error from
systematic variations in the large-scale flat-field is 0.033\%,
equivalent to 100 $\rm e^{-} pixel^{-1}$  of sky background.

{\it Intrinsic variation in galaxy brightness}

Following MBH (\cite{MBH94}) and Zheng et al., we can calculate the
random error due to intrinsic variation in the surface brightness of
the galaxy. In the terminology introduced by Tonry and Schneider (1988), 
m1 for our image is 20.30 mag, the exposure time is 154,043 sec, the
distance is 14.5 Mpc and $\rm \bar{M}_{6660}$ is adopted as 0
(\cite{Zheng99}). In the case of 300 $\rm e^- pixel^{-1}$ from the galaxy
(equal to a surface brightness of 28.23 mag $\rm arcsec^{-2}$), this
error is 53.8 $\rm e^{-} pixel^{-1}$, or an error of 18
translates to an average error of 1.29 $\rm e^- $ for the 
largest bins used here, to 10.98 $\rm e^- $ for the smallest bins.

{\it Background subtraction}

There are two types of background subtraction errors, both dominated
by systematic effects.  One is the accuracy of the sky background
subtraction, for which we adopt the value of 115 e$^{-}$ obtained
above.  The second comes from imperfect star subtraction, especially
for saturated stars.  For this we adopt an average error of 100
e$^{-}$/pixel by checking the regions in which stars are subtracted.
We acknowledge this error can be significantly higher for selected
stars, especially for the saturated stars which are so close to
NGC~4565.  For the true errors involved in star subtraction, we must
look to the consistency of the luminosity profiles at faint surface
brightness levels.

\subsection{The Total Error Budget}

All the sources of errors discussed are listed in Table 2.  We assume
that the mean count per pixel from the object galaxy is 300 e$^{-}$
(i.e., in its faint halo), a sky level of 302,500 e$^{-}$ and bin 
sizes of 35$\times$50 pixels and 8$\times$3 pixels for surface 
photometry. This leads to a final error of close to 400 $\rm e^{-} 
pixel^{-1}$.  Errors for each bin are calulated based on the number 
of non-masked pixels in that bin, ratioed to the error expected for 
all pixels in that bin having data.

As photon noise in the sky is the dominant source of error in our
data, we give the expected error per bin for both the smallest bin we
use for $R$-profiles (8$\times$3 pixels) and the largest bin we use
for $z$-profiles (35$\times$50 pixels).  The error for the faintest
part of NGC~4565 is close to 183 $\rm e^{-} pixel^{-1}$, or 63 $\rm
e^{-} arcsec^{-2}$. That is, the relative error of measured flux at
28.77 mag arcsec$^{-2}$ is 100\%, leading to an error bar of 0.75 mag
arcsec$^{-2}$. This also corresponds to an error of 0.25 mag
arcsec$^{-2}$ at a surface brightness of 27.5 mag arcsec$^{-2}$. In
fact, the error is a little higher than this, because there exist
masked regions in the measured boxes. The main sources of error are
from large-scale variation of flat-field, sky fitting and residuals
from star subtraction.

\begin{table}[]
\caption[]{The error estimation}
\vspace {0.5cm}
\begin{tabular}{lrrrrrr}
\hline
\hline
Source of variation & $\rm e^{-}/pixel$ & \% & 35$\times$50
bin(e$^-$)$^a$ &\% &8$\times$3 bin(e$^-$)$^a$ & \% \\
\hline
Readout noise & 147.0 & 0.049 & 3.5 & 0.001 & 30.0 & 0.010\\
Photon noise of sky & 320.0 & 0.106 & 15.0 & 0.005 & 130.6 & 0.043\\
Dark & 11.3 & 0.004 & 0.27 & 0.0001&2.3 & 0.0008\\
Bias (formal) & 0.8 & 0.0003 & 0.02& 0.0000&0.17 & 0.0001 \\
Flat field (small-scale/pixel scale)& 26.9 & 0.009 & 0.6 & 0.0002& 5.1
& 0.0017\\
Flat field (large-scale/$>100$ pixel scale)& 100.0 & 0.033 & 100.0 &
0.033 & 100.0 & 0.033\\
Surface brightness fluctuations$^b$ & 53.8 & 0.018 & 1.29 & 0.0004 &
10.98 & 0.0036\\
Background subtraction & 115.0 & 0.038 & 115.0 & 0.038& 115.0 &
0.038\\
Star subtraction & 100.0 & 0.033 & 100.0 & 0.033 & 100.0 & 0.033\\
                                &         &         &       &  & & \\
Total & 401.2 & 0.133 & 182.9 & 0.060 & 226.6 & 0.075\\
\hline
\end{tabular}\\
$^a$ The error in surface photometry for bins of $\rm 35\times50$ or
$\rm 8\times3$ at a count of sky level\\
$^b$  Here assuming 300 $\rm e^- pixel^{-1}$ from the galaxy.
\end{table}

\section{ANALYSIS AND RESULTS}

\subsection{Luminosity profiles}

As others have done (cf. NJ), we use a contour map to determine the
position angle of the major axis.  Then the sky-subtracted image is
rotated to that angle to put the major axis of the object galaxy along
the x-axis direction. At the same time, all circular masks but the
three masks used for the object galaxy were transformed to the new 
coordinate system and imposed on the rotated image (Figure~4).

\begin{figure}
\figurenum{4}
\caption[]{This is the image that results from sky--subtracting
a two-dimensional background from the final image, with the final
point source masks still in place.  We have rotated the image so that
the major axis of the galaxy lies along the x-axis.
Here we only present the central $\rm 700\times700$ pixels
region of the image.
\label{fig4}}
\end{figure}

For purposes of analysis, we measure the luminosity profiles of
NGC~4565 in two orthogonal directions. One direction is parallel to
the galaxy minor axis (the ``$z$'' direction), the other is parallel to
the major axis (the ``$R$'' direction).  We sample the $z$ direction in
discrete lengths (parallel to the major axis) of 35 pixels ($\sim$1 arcmin) 
on both sides of the major axis.  We employ bins of varying width 
perpendicular to the major axis (i.e., along the $z$ direction): 
from 1 pixel when $z$ = 0 (i.e., on the major axis) to 50 pixels for 
the largest $z$ distances ($\sim 7'$).  This methodology yields bin 
sizes that vary from 35$\times$1 pixels to 35$\times$50 pixels.  
Only unmasked pixels are used, and the flux cited is the median value 
for that bin.  The measured $z$ and $R$ profiles are shown in
Figure~5. Here we adopt a distance of 14.5 Mpc, with $1'' = 0.0703$ kpc.

We see that the four $z$-profiles in each plot (i.e., all four quadrants
of the galaxy) agree reasonably well down to surface brightness of 28
mag arcsec$^{-2}$, or a level of 0.12\% of sky) within a radius of 6
arcmin.  Greater than that distance, the well-known warp in the disk
of this galaxy (SE to NW) (cf. \cite{S76}; \cite{K79}; \cite{NJ97};
see below) begins to break down the symmetry.  The $z$-profiles are
tabulated in Table 3 (full table in electronic form).

The $R$ series of profiles are parallel to the major axis with $z$
distance: 0$''$,10$''$,20$''$,30$''$,40$''$,60$''$,80$''$,100$''$ and
120$''$ respectively. Since the $z$-profiles decreases quickly at 
small $z$ and relatively slow at larger $z$, the sizes of rectangles are
changed from 8$\times$3 pixels at $z = 0''$ to 8$\times$11 pixels
at $z = 120''$.  The measured $R$ profiles are also in Figure~5.  The
asymmetries of the NW and SE parts are also shown in $R$-plots. The data
are listed in Table 4.

\begin{figure}
\figurenum{5}
\vspace{-0.9cm}
\caption[]{The $z$-profiles (upper plots) and $R$-profiles (lower
plots) of NGC~4565 from our 6660$\rm\AA$ final image.  The four
quadrant profiles of the galaxy (except the minor and major axes)
are plotted in each sub-plots. The $z$-profiles agree reasonably well
down to 28 mag within a radius of 6 arcmin.  The obvious asymmetries
of the NW and SE parts are shown in both $z$-plots and $R$-plots. The dust
lane is indicated by arrows in the first 3 $z$-plots. The solid line 
is our two-disk + halo model. The dash-dot and dashed lines are the 
thin and the thick disk models respectively, and the dotted line is 
the power-law halo. All the data are well--fitted by this model.  
Here one arcsec is 0.0703 kpc.
\label{fig5}}
\end{figure}

\begin{table}[]
\caption[]{Surface brightness of $z$-profiles of NGC 4565 in i(6660$\rm \AA$)-band }
\vspace {0.5cm}
\begin{tabular}{lccccc}
\hline
\hline
 $R$ & $z$ & $z$ & i(6660$\rm \AA$)& High Error& Low Error \\
 (arcmin) & (arcsec) & (kpc) & ($\rm mag/arcsec^2$)&($\rm mag/arcsec^2$) & ($\rm mag/arcsec^2$) \\
\hline
 0.00 &      2.57 &    0.18 &  19.50 &    0.04 &      0.04\\
      &      5.99 &    0.42 &  20.27 &    0.03 &      0.03\\
      &      9.41 &    0.66 &  20.59 &    0.02 &      0.02\\
      &     12.83 &    0.90 &  20.37 &    0.02 &      0.02\\
      &     16.25 &    1.14 &  20.07 &    0.02 &      0.02\\
      &     19.67 &    1.38 &  19.85 &    0.01 &      0.01\\
      &     23.09 &    1.62 &  19.93 &    0.01 &      0.01\\
      &     26.51 &    1.86 &  20.16 &    0.01 &      0.01\\
      &     29.93 &    2.10 &  20.42 &    0.02 &      0.02\\
      &     33.35 &    2.34 &  20.73 &    0.02 &      0.02\\
\hline
\end{tabular}\\
{Note.-- Only a portion of Table 3 listed here.}
\end{table}

\begin{table}[]
\caption[]{Surface brightness of $R$-profiles of NGC 4565 in the i(6660$\rm\AA$)-band}
\vspace {0.5cm}
\begin{tabular}{lccccc}
\hline
\hline
 $z$ & $R$ & $R$ & i(6660$\rm \AA$)& High Error& Low Error \\
 (arcsec) & (arcsec) & (kpc) & ($\rm mag/arcsec^2$)&($\rm mag/arcsec^2$) & ($\rm mag/arcsec^2$) \\
\hline
  0.00   &   0.00  &  0.00  &  17.55   &   0.10  &     0.11\\
         &  10.26  &  0.72  &  18.45   &   0.09  &     0.09\\
         &  20.52  &  1.44  &  18.85   &   0.06  &     0.06\\
         &  30.78  &  2.16  &  19.23   &   0.07  &     0.08\\
         &  41.04  &  2.89  &  19.57   &   0.07  &     0.07\\
         &  51.30  &  3.61  &  19.74   &   0.06  &     0.06\\
         &  61.56  &  4.33  &  20.01   &   0.07  &     0.07\\
         &  71.82  &  5.05  &  20.22   &   0.06  &     0.07\\
         &  82.08  &  5.77  &  20.53   &   0.02  &     0.02\\
         &  92.34  &  6.49  &  20.53   &   0.02  &     0.03\\
\hline
\end{tabular}\\
{Note.-- Only a portion of Table 4 listed here.}
\end{table}

\subsection{Warp}

Figure~6 shows the central points of the profiles parallel to the
minor axis. This was determined by keeping the upper and lower
profiles in the best agreement.  By plotting these central points as a
function of radius in this galaxy, we map out its warp from SE to NW
sides. The NW warp was first found in HI (\cite{S76}) and the
optical warp of both sides was discussed by NJ.

The figure shows that the stellar warp starts at 21 kpc from the
center of galaxy and is less distorted than that of NGC 5907
(\cite{Zheng99}; \cite{MBH94}), whose warp starts from 4.1 kpc.  In
fact, the starting points of warp are asymmetric: 21 kpc NW and 25 kpc
SE.  This causes the asymmetries in $z$-plots and $R$-plots (Figure
5). The amplitude of the warp increases rapidly after it begins,
becoming more than 1 kpc from the centrally--defined major axis after
a distance of 30 kpc.  We plan to take further deep images of NGC~4565
to better explore its fainter stellar populations, including those
in its warp.

\begin{figure}
\figurenum{6}
\centerline{\psfig{file=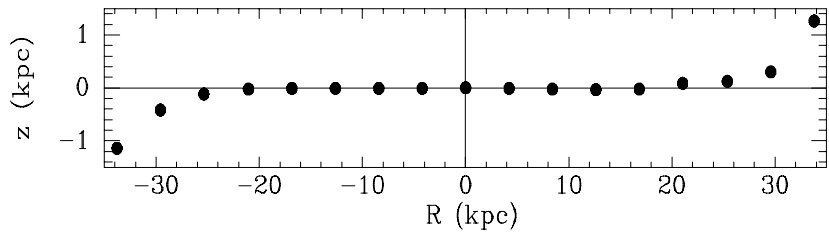,width=16.0cm}}
\caption[]{The stellar warp of NGC 4565 measured in 6660$\rm
\AA$ band.  The x-axis is along the major-axis of galaxy. The left is 
to the south-east (SE) and the right is to the north-west (NW). The
warp can be seen on both sides of this galaxy, but starting at
different distances on either side.
\label{fig6}}
\end{figure}

\subsection{Comparison with previous work}

Jensen and Thuan (1982) present the luminosity profile of NGC~4565 in
the B and R bands, using photographic data.  As our intermediate-band
i filter centered at 6660$\rm\AA$ is quite similar to R-band (cf. Fan
et al. 1996), it is straightforward to compare with JT's results.
Zhou et al. (2001) give the transformation between the BATC i-band and
R-band as: $\rm R = m_{6660} + 0.1048$.  The data of JT are then
compared to our own as a function of the $z$ direction for six values of
$R$, as shown in Figure~7.  Here we plot the difference between observed
profiles and our fitted profile , as a function of $z$ distance.  In
general, the JT's data (upper plots) agree with ours (lower plots)
within 0.5 mag, but the data of JT become systematically
fainter than ours at large radii and $z$-distances.  Given that the data
of JT are photographic, their faint values are more subject to larger 
systematic problems than ours.

\begin{figure}
\figurenum{7}
\centerline{\psfig{file=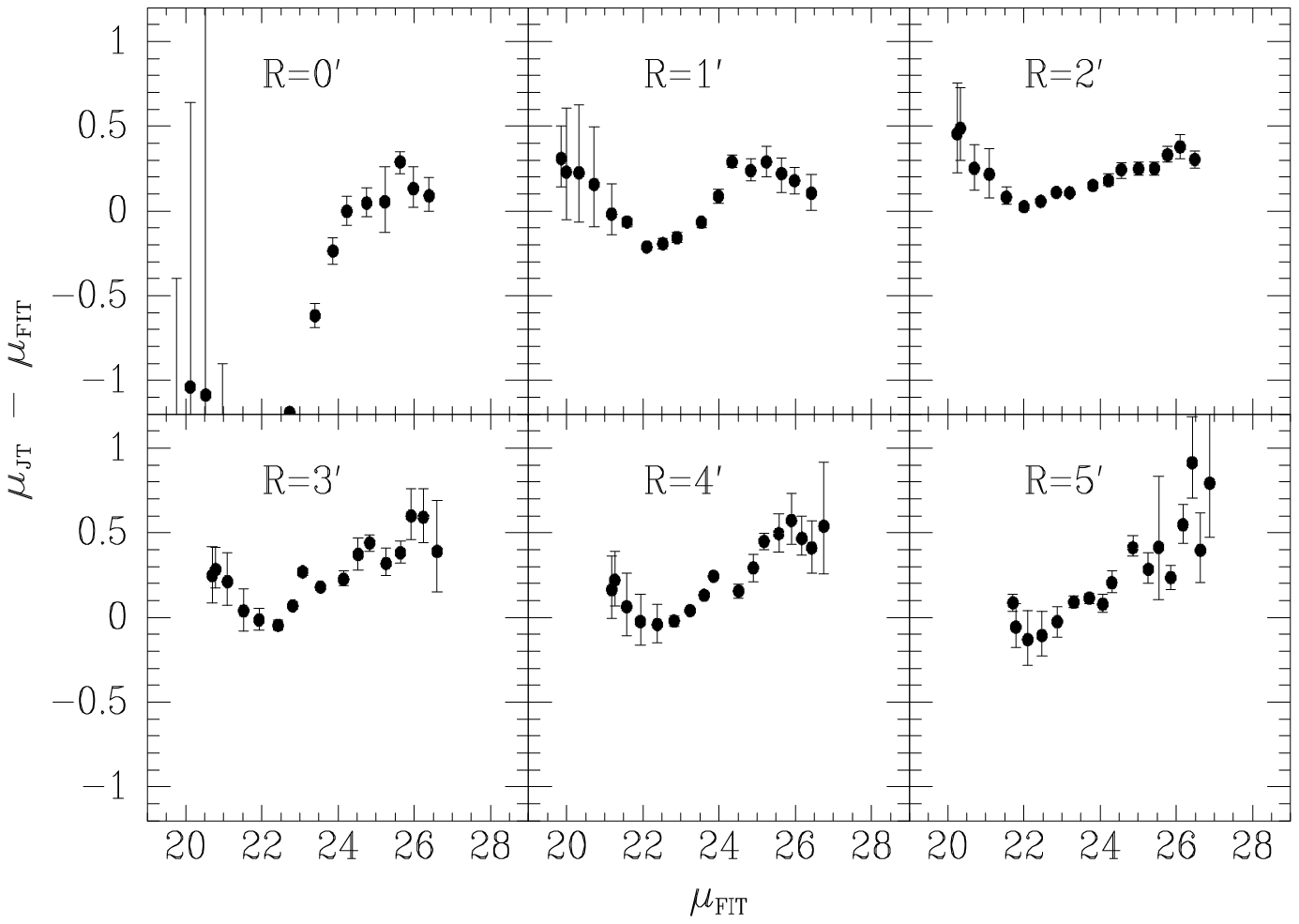,width=16.0cm}}
\centerline{\psfig{file=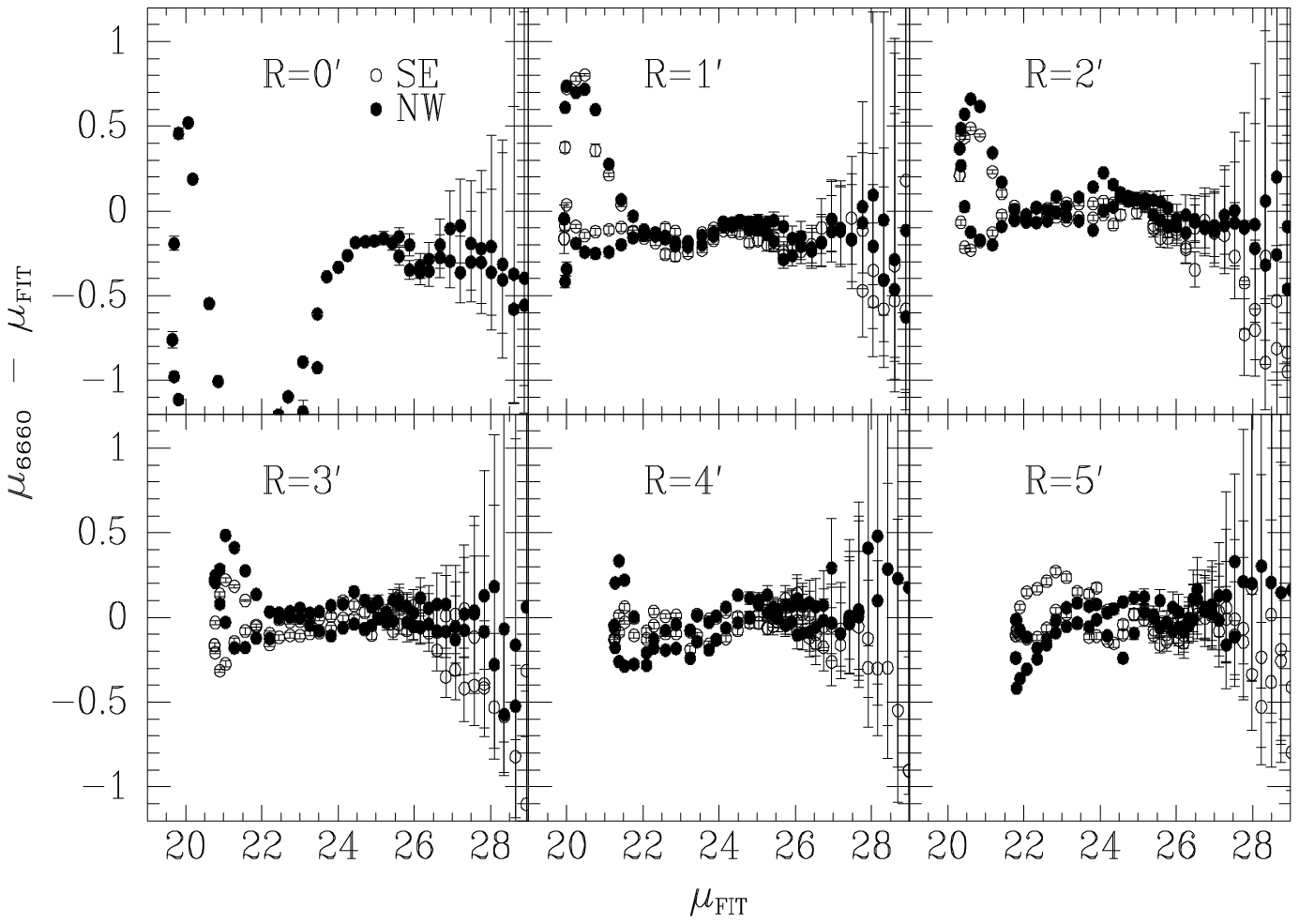,width=16.0cm}}
\caption[]{Comparison of our $\rm 6660\AA$ data with those of
JT. Differences between observed profiles (JT, upper plots; and ours,
lower plots) and our fitted profiles are plotted against model surface
brightness.  In the lower plots, the solid circles are data from the 
NW side of the galaxy and open circles are data from the SE side.
\label{fig7}}
\end{figure}

\subsection{The adopted model}

Previous studies have shown that one single disk can not fit the
fainter parts of the $z$-profiles of NGC~4565 reasonably (\cite{JT82};
\cite{SG89}; \cite{NJ97}).  Rather, for the $z$ profiles we adopt a
three-component model: thin disk (seen at inclination $i$), thick disk
(cf. \cite{Burstein79}), and halo.  The $z$ structure for the thin and
thick disks adopted is that of the sech$^{2}(z)$ model (van der Kruit
\& Searle 1981), which has a density distribution of a
self-gravitating isothermal sheet in the $z$ direction.  Alternate
fits proposed by \cite{GK96} (i.e., exponential or sech($z$) light
distributions), while fitting better at small $z$ values, are not as
physically motivated as the $\rm sech^2$ model.  Moreover, all three
models show the similar behavior at large $z$, showing that the
different $z$ models do not affect the model of the halo one derives
for edge-on galaxies.  A standard exponential disk was used in the
radial direction ($r$), well away from the bulge.

The density distribution of both the thin and the thick disks is
described by

\beq
\rho(r,z)=\rho_0 e^{-\frac{r}{h_r}} {\rm sech}^{2}(\frac{z}{z_0}),
\eeq

\noindent with $z_0$ and $h_r$ as the disk scale height and scale
length and $z_0$ equal to twice the scale height $h_z$ of the
exponential model.  As galaxy disks do not go to infinity in either
the $z$ or the $r$ directions (cf. Figure~4), we introduce a cutoff in
the $r$-distribution at $r_{max}$, beyond which for $ r>r_{max}$,
$\rho(r,z)=0$.  For an edge-on galaxy, this means

\beq
\mu(R,z) = \mu_0 - 2.5 \log10\left[2 {\rm sech}^{2}(\frac{z}{z_0})
\int_{R}^{r_{max}}\frac{e^{-\frac{r}{h_r}} r}{\sqrt{r^2-R^2}} \:
dr\right].
\eeq

\noindent Here, $R$ is the projected distance from the center along
the major axis.

A power-law halo is introduced as the third component, with its
luminosity distribution modeled as in NJ:

\beq
\rho(r,z)=\frac{\rho_0} 
{[1+\frac{r^2+(\frac{z}{q})^2}{r_0^2}]^{\frac{\gamma}{2}}},
\eeq

\noindent where $r_0$ is the core radius, $q$ is the axis ratio and
$\gamma$ is the power index. For an edge-on galaxy, this model has the
form (where $\Gamma$ is the standard Gamma function):

\beq
\mu(R,z) = \mu_0 - 2.5 \log10\left\{{\sqrt{\pi}} \frac{\Gamma
(\frac{\gamma -1}{2})}{\Gamma (\frac{\gamma}{2})} r_0^{\gamma}
\left[r_0^2+R^2+(\frac{z}{q})^2\right]^{\frac{1-\gamma}{2}}\right\}.
\eeq

\subsection{Profile fitting}

\subsubsection{Data used for fitting}

As its dust lane nearly bisects the major axis of NGC~4565, it is
difficult for us to fit all the components of this galaxy (nucleus,
bulge, disks and halo) simultaneously.  The primary purpose of the
present paper is to search for luminous halos in edge-on
galaxies. Hence, we concentrate only on the three components
whose luminosity distributions affect the halo of this galaxy: thin
disk (via projected major axis), thick disk and halo.  To avoid the
possible effect of nucleus, bulge, dust lane and warps, only the
$z$-profiles with 2$'$ $\leq R \leq$ 7$'$ were used in the fit.  For
$R$-profiles, we rejected the profiles along the dust lane and the
points within bulge.  Though there exists obvious asymmetries between
the NW and SE parts of galaxy, they are treated here the same owing to
the fact that large areas in both sides of the galaxy are masked.

A general $\chi^{2}$ method was used to estimate the parameters
of the fits.  In principle, one should use a weighted fitting 
scheme (cf. Shaw \& Gilmore 1989) of the form:

\beq
\chi^{2} = {\sum}_i[w_i(\mu_{i}-\mu_{fit})]^2,
\eeq

\noindent with $\mu_i$ and $\mu_{fit}$ as observed and fitted surface
brightnesses.  In practice, the values assigned to $w_i$ depend on
what one is trying to fit.  Unfortunately, the manner in which we can
compensate for fainter surface brightness levels (by using larger
sample bins) does not fully compensate for the observational errors in
the outer regions, which are much larger than those in the inner
regions.  In addition, by sampling with different bin sizes, we have
more measurements nearest the major axis than we do far away from the
major axis.  Thus, we have more points in our fit for the inside
regions with small uncertainties than those for the outside with
large uncertainties.  Strictly error--weighting the data would bias the
fits toward the innermost data.  As such, we chose to give equal
weight to the bins, which better accounts for the larger number of
bins close to the major axis.  The fitting algorithm was implemented
using the Levenberg-Marquardt technique, with the subroutine used
taken from Numerical Recipes (\cite{Press92}).

\subsubsection{Estimates of parameters}

Three parameters must be determined before fitting: the coordinates of
the galaxy center $x_c$ and $y_c$ and the sky background value.  The
latter component has already been removed from the data to be
analyzed.  The effect of inclination is neglected here (but considered
later), as NGC~4565 is so nearly edge-on.  This leaves us 12
parameters to fit, namely: 3 surface brightness $\mu_{01}$, $\mu_{02}$
and $\mu_{03}$ (with 1 = thin disk; 2 = thick disk and 3 = halo), the
scale heights $z_{01}$ and $z_{02}$ of thin and thick disks components
in $z$-direction and scale lengths $h_{r1}$ and $h_{r2}$ in radial
direction, the cutoff radius $r_{max1}$ and $r_{max2}$ of two kinds of
disks, and the core radius $r_0$, axis ratio q and power index
$\gamma$ of the halo.

Unfortunately, we have too many parameters for the fitting procedure
to converge successfully.  As discussed by Fry et al. (1999), for three
free parameters, the initial estimates need to be within 20\% of the
correct values for the model to converge. At the same time, the
existence of many local minima make the final fit strongly dependent
on the initial estimate of these parameters.

To find the true minimum and the best fit, we randomly selected
10,000 sets of 12 initial parameters spaced uniformly as to cover the
reasonable range covered by these parameters (Table 5).  Of these
10,000 tries, we found nearly 3,700 convergences.  The $\chi^2$ for
these 12 parameters from these 3,700 convergences are plotted against
the value for finding for those convergences in Figure~8.  The minimum
is marked as symbol star in each sub-plot.

\begin{table}[]
\caption[]{The confined ranges of initial parameters.}
\vspace {0.5cm}
\begin{tabular}{lll}
\hline
\hline
 Thin disk  & Thick disk &  Halo(power-law)   \\
\hline
$\rm\mu_0$: 20 - 23 mag $\rm arcsec^{-2}$ & $\rm\mu_0$: 23 - 27 mag
$\rm arcsec^{-2}$ & $\mu_0$: 22 - 27 mag $\rm arcsec^{-2}$ \\ 
$z_0$: 0.6 - 1.5 kpc & $z_0$: 1.2 - 2.9 kpc & $r_0$: 0 - 22 kpc\\
$h_r$: 5.8 - 11.6 kpc & $h_r$: 7.2 - 29.0 kpc & $\rm\gamma$: 2 - 5 \\ 
$r_{max}$: 29 - 38 kpc & $r_{max}$: 32 - 39 kpc & $q$: 0.1 - 0.9 \\
\hline
\end{tabular}\\
\end{table}

The distributions of $\mu_{01}$, $z_{01}$, $h_{r1}$ (i.e., the thin
disk component) and $q$ (the axis ratio of the halo) are well determined
by our data.  The axis ratio of the halo we obtain (0.44) is in good
agreement with that of NJ, who show that $q$ = 0.5 is a better fit
than $q$ = 1.0 for a two-disk-plus-halo model in the V-band.

The parameters of the thick disk and the power-law halo have wide
$\chi^2$ distributions. The scale height of the thick disk (2.55 kpc)
is about twice of that of the thin disk and agrees with the $h_z=1$
kpc ($h_z=z_0/2$) given by \cite{Burstein79} for S0 galaxies.  The
thick disk scale length is 11.03 kpc, larger than that of the thin
disk.  The luminosity of the thick disk is 20\% of the thin disk. The
core radius of the power-law halo is 14.4 kpc, indicating a flat
halo. This agrees with the third component of three disk models of NJ
and that of Shaw \& Gilmore (1989), in which the scale lengths of
their third disks are about 13.9 kpc (at a distance of 14.5 Mpc).  The
values of $\gamma$ concentrate in the range of 3.2 to 4.0, which
indicates that there does not exist a $r^{-2}$ halo in this galaxy. 
The best power index is 3.88, between the 3.5 of Milky Way 
(\cite{Zinn85}) and the 4.0 of M31 (\cite{PB94}).  The $r_{max}$ of
the thin and the thick disks show several local minimums, owing to 
data sampling. The cutoffs of the thin and thick disks are about 
32 kpc and 37 kpc, respectively.

Analyses of correlation between any two parameters show that some
of the parameters are correlated, which means some of the parameters
can not be determined independently.  For example, we find correlation
between $z_{01}$ and $z_{02}$ and among $\mu_{03}$, $r_0$ and
$\gamma$. The similar $\chi^2$ distributions of $ \mu_{03}$, $ r_0$
and $ \gamma$ in Figure~8 also show such a correlation.

The error for each parameter is obtained in the following way.  
Random values are selected for the observed data such that they obey
a normal distribution, with sigmas determined by the known errors in
each sampled bin.  We then obtain the best--fitted parameters for that
set of data. This procedure is repeated three hundred times, giving
us 300 separate determinations of the best-fitted value for each
parameter.  The statistical standard deviation of each parameter 
from this procedure is adopted as the final error for this parameter.

The best fit parameters and errors are listed in Table 6. Table 6 also
lists the total magnitude in 6660$\rm\AA$-band to a surface brightness
of 28 mag arcsec$^{-2}$, $\rm m_{6660} = 8.99$ (or broad-band R = 9.10 
from the Zhou et al. transformation), which is measured by replacing
masked areas around the galaxy by the corrsponding parts of the
unmasked galaxy.

\begin{figure}
\figurenum{8}
\caption[]{The values of the $\chi^2$ for the 3,700 random fits
that converged, versus the value of the parameters at which they
converged. The star symbol in each sub-plot is the minimum point
adopted as our best-fitting value.  The values of $\gamma$ concentrate
in the range of 3.2 and 4.0, ruling out the existence of luminous halo
in NGC~4565.
\label{fig8}}
\end{figure}

\begin{table}[]
\caption[]{ parameters of the thin-disk + thick-disk + power-law halo
model.}
\vspace {0.5cm}
\begin{tabular}{lll}
\hline
\hline
 Thin disk  & Thick disk &  Power-law halo    \\
\hline
$\mu_0=22.32^{-0.04}_{+0.08}$ mag $\rm arcsec^{-2}$ &
$\mu_0=25.52^{-0.50}_{+0.41}$ mag $\rm arcsec^{-2}$ &
$\mu_0=26.86^{-0.64}_{+0.86}$ mag $\rm arcsec^{-2}$ \\
$z_0=1.17^{-0.04}_{+0.04}$ kpc & $z_0=2.55^{-0.30}_{+0.22}$ kpc &
$r_0=14.44^{-4.07}_{+4.74}$ kpc \\
$h_r=8.05^{-0.19}_{+0.34}$ kpc & $h_r=11.03^{-1.88}_{+1.12}$ kpc &
$\gamma=3.88^{-0.31}_{+0.56}$ \\
$r_{max}=31.84^{-0.12}_{+0.07}$ kpc & $r_{max}=36.93^{-0.64}_{+0.22}$
kpc & $ q=0.44^{-0.02}_{+0.02}$ \\
$(m_{6660})^a=8.99\pm0.02$ mag & $(L_{thick}/L_{thin})^b=20\%$&
$(L_{halo}/L_{thin})^c=21.4\%$ \\
\hline
\end{tabular}\\
$^a$- Total photometrical magnitude of NGC~4565 at 6660$\rm \AA$ to
surface brightness of 28 mag $\rm arcsec^{-2}$.\\
$^b$- The luminosity ratio of model thick-disk and thin disk.\\
$^c$- The luminosity ratio of model power-law halo and thin disk.\\
\end{table}

Figure~5 presents the best fitting values for the three components 
compared to our data in both the $z$ and $R$ directions.  The fits are
quite good in the range of $R = 1'$ to $R = 8'$ for the
$z$-profiles and in the range of $z= 30"$ to $120"$ for $R$-profiles.  As
expected, the model deviates significantly from the data in the region
of the bulge (not-fitted) and where the warp of the disk becomes sizeable.

Table 7 compares the dimensional parameters obtained by this study and
to similar parameters determined by other studies of NGC~4565, as
magnitude zero points are too dependent on the filter that was used.
The thin disk parameters are consistent among the different analyses.
Our value of 8.05 kpc for the thin disk exponential scale length and
1.17 kpc for its scale height are close to those obtained by R96, KS, NJ
and SG, scaled to our adopted galaxy distance.  The scale height and
scale length of our fits to the thick disk agree best with the thick
disk in SG's two--disk + R$^{1/4}$ halo model. The thick disk
components of our model are somewhat larger than the thick disk in the
three--disk models, likely owing to the halo being much flatter in the
inner region of NGC~4565 than the models for the third disk used by
these other studies.

\begin{table}[]
\caption[]{A comparison of the parameters obtained from different
sources.}
\vspace {0.5cm}
\begin{tabular}{lcccccccc}
\hline
\hline
component &filter & thin-disk& & thick-disk & & third-disk(exp) & or
halo & \\
& & $h_r$ (kpc) & $z_0$ (kpc) & $h_r$ (kpc)& $z_0$ (kpc) &
$h_z$ (kpc) & $q$ & $\gamma$ \\
\hline
KS          &$\rm B_J$ &  8.0& 1.15  &      &   &  &  & \\
H80         &B&  8.8&      &      &         &   & 0.46& 3.46 \\
H82         &B&  9.7 &     &      &         &  & 0.50& 4.10 \\
JT          &B&  9.9 &     &  9.9& 1.88 &    4.78& 0.52 & \\
R96         &K&  8.4 &1.28$^a$ &     &     &       & & \\
SG$^b$ &B&  10.6 &1.10 & 6.2 &1.62 &   3.42& & \\
      &R&  6.8 & 1.07&  7.5& 1.90&    4.23& & \\
NJ$^b$    &V&  8.1 &0.87 & 8.6 &1.88 &      5.1& & \\
SG$^c$&B&  8.0 &1.20 &12.9 &2.28 &   &0.37 & \\
 &R&  7.0 &1.13 & 9.1 &2.29 &   &   0.36 & \\
KB & V &  & && &   & &3.46 \\
SOSCB & $\rm 6040\AA$ &  & && &   & &3.78 \\
This work$^d$ &$\rm 6660\AA$ & 8.05&1.17 &11.03&2.55 & &0.44 &3.88 \\
\hline
\end{tabular}\\
$^a$ - R96 use sech disk model give $h_z$=0.44 which corresponding to
$z_0/2$.\\
$^b$ - sech$^2$+sech$^2$+exp model\\
$^c$ - sech$^2$+sech$^2$+$\rm R^{1/4}$ halo model\\
$^d$ - sech$^2$+sech$^2$+power-law halo model\\
Note.- All the values in the table are transformed to the distance of
14.5 Mpc.\\
KS: \cite{KS81}; H80: \cite{H80}; H82: \cite{H82}; JT: \cite{JT82};
R96: \cite{R96}; SG: \cite{SG89}; NJ: \cite{NJ97}; KB: \cite{KB78};
SOSCB: \cite{SOSCB} \\
\end{table}

\subsection{Bright Versus Faint Parts of NGC~4565}

As our data go very faint, one worry is that the very faintest parts
of the PSFs of the bright stars could interfere with our measurement
of the halo of this galaxy (such as we found to be the case in our
analysis of NGC~5907; cf. Zheng et al.).  As our final image of
NGC~4565 did not permit us to fully test the faintest parts of the PSF
in a reliable manner, we separately observed a set of bright stars
with the same CCD parameters. From those observations we construct a
new PSF that can extend to 1700 arcsec, compared to the extension of
only 510 arcsec permitted by our final NGC~4565 image.  (In this we
note that even 510 arcsec is still farther than the radii considered
by either Morrison et al.(1994) for NGC 5907 or Fry et al.(1999) for
NGC 4244).  The agreement between the PSF we obtain from our NGC~4565
image and that obtained from the special bright star images is
excellent where they overlap (i.e., interior to 510 arcsec).

There are two effects that can, in principle, affect our halo results.
One is scattered light from the nucleus and bulge. The other is the
scattered light from the brighter parts of the galaxy as a whole.  To
test the first effect, the nucleus and bulge is separated from the
other parts of galaxy by subtracting our best fitting model from the
final image (cf. Figure~11).  We then convolve the resulting nucleus
and bulge with the extended PSF.  The result is shown in Figure~9.
The $z$-profiles of our model are represented by solid lines, and the
outer wings due to scattered light from the nucleus and bulge are
represented by dashed lines.  We find that scattered light from the
central region of NGC~4565 at least 2 mag arcsec$^{-2}$ fainter than
the model fit everywhere, including on the minor axis.

\begin{figure}
\figurenum{9}
\centerline{\psfig{file=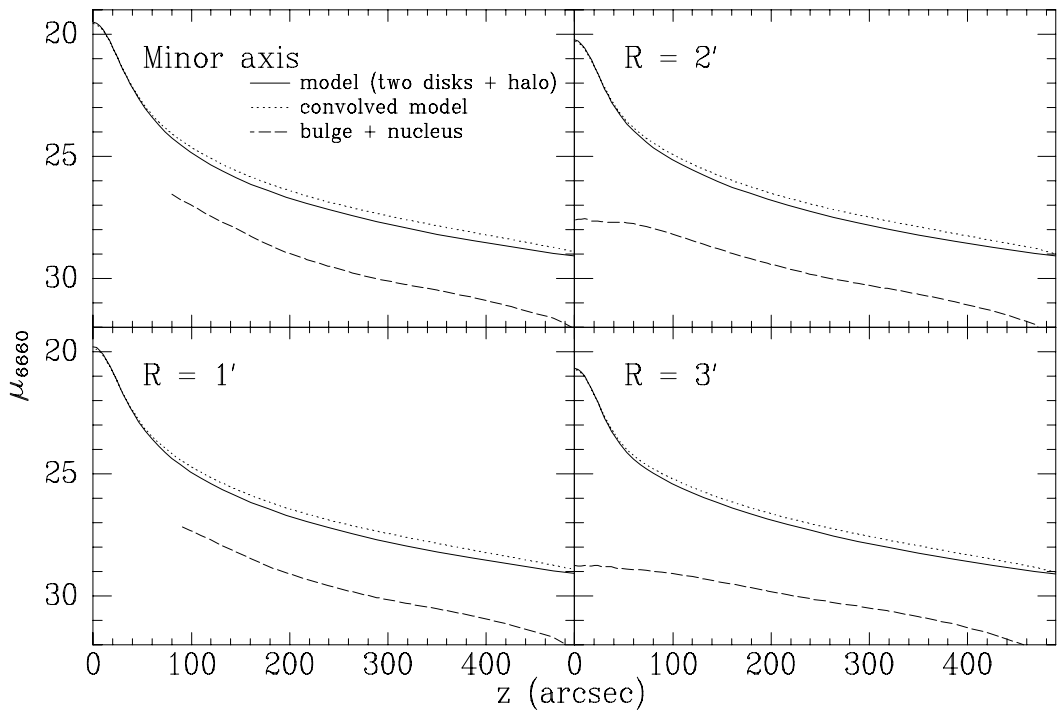,width=16.0cm}}
 \caption[]{The possible effects of scattered light from the
brighter parts of NGC~4565 on the model obtained from its faintest
parts.  The solid lines are our model, the dashed lines represent
the scattered light from the nucleus and bulges, and the dotted lines
are the model convolved with extended PSF.
\label{fig9}}
\end{figure}

To test the effect of scattered light from the galaxy as a whole,
we convolved the model with the full PSF (dotted line in Figure~9).
As is evident, the extended PSF has little effect on the model
we derive for the faintest parts of NGC~4565, being at most
0.2 - 0.3 mag different at very faint surface brightness, where
the errors of measurement are at least that high.  We conclude
that, within the errors of measurement, our parameters for the
halo of NGC~4565 are not significantly affected by scattered light 
from the brighter parts of the galaxy.

\subsection{Inclination}

We estimate the inclination of the disk of NGC~4565 by assuming that
the thin disk of this galaxy is round with a radius of 38 kpc, and the
dust lane is flat and lies at the mid-plane of the disk.  We assume
that the upper-edge (cf. Figure~4) of the dust-lane could be the edge
of round disk.  By measuring the projected distance between the
upper-edge of the dust-lane and the major axis of NGC~4565 in the
plane of the sky, taking into account the radius of the disk, we
obtain an inclination of 87.5$\rm ^{\circ}$, which is somewhat higher
than that of 86$\rm ^{\circ}$ in KS(1981).

Given our model assumes that the galaxy is exactly edge-on ($i = 
90^\circ$), we can test how our model changes for different values
of $i$ via the following transformation:

\beq
\sigma(R,z) = \int_{\frac{-\sqrt{r_{max}^2-R^2}-z 
\cos{i}}{\sin{i}}}^{\frac{\sqrt{r_{max}^2-R^2}-z \cos{i}}{\sin{i}}} 
{\rm sech}^2(\frac{\mid z\sin{i}-l\cos{i} \mid}{z_0}) 
\exp({\frac{-\sqrt{R^2+(z\cos{i}+l\sin{i})^2}}{h_r}}) dl .
\eeq

\noindent The variable of integration, $l$, is along the
line-of-sight.  We plot the minor axis profiles with 90$^{\rm \circ}$,
87.5$\rm ^{\circ}$, and 86$\rm ^{\circ}$ for both the thin and the
thick disks in Figure~10.  As might be expected, the form of the thick
disk is relatively insensitive to the exact inclination used for this
galaxy.  Even the thin disk model is only slightly more sensitive,
having a central surface brightness $\sim 0.1-0.2$ mag arcsec$^{-2}$
fainter for 87.5$\rm ^{\circ}$ than for $\rm 90^\circ$.

\begin{figure}
\figurenum{10}
\centerline{\psfig{file=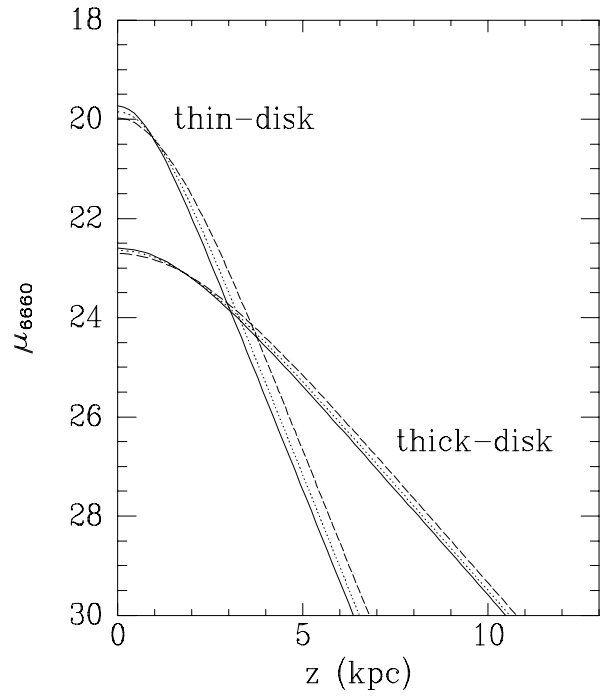,width=16.0cm}}
\caption[]{The effect of inclination on the model
profiles of the minor axis of the thin and thick disks.  Three
inclinations are considered: 90$\rm ^{\circ}$(solid lines), 87.5$\rm
^{\circ}$(dotted lines) and 86$\rm ^{\circ}$(dashed lines).
\label{fig10}}
\end{figure}

\subsection{Bulge}

The bulge can be obtained, after subtraction of the disks and halo.
We show the two-dimensional distribution of the bulge and the profile
along the minor axis in Figure~11.  The bulge extends up to 5 kpc, and
it is clear that it cannot be described by R$^{1/4}$--law (dashed
line) from 0.7 kpc to 5 kpc.  Rather, an exponential distribution
(cf. \cite{KB78}) (solid line) with a scale height of $h_z = 0.65$ kpc
($\sim z_0/2$) is the best fit to the bulge luminosity distribution,
consistent with the values of 0.72 kpc determined by JT and 0.75 kpc
in the K-band (\cite{R96}).  We find the bulge to have an ellipticity
of 0.8, based on a simple fit using elliptical isophotes.

Speculation abounds in the literature about the nature of the bulge
seen in NGC~4565.  Is it bar-shaped, seen with the long axis projected
along our line-of-sight (much like the bar of our own galaxy; cf.
\cite{Blitz91})?  Or is it dynamically coupled to the thin
disk (which has a similar exponential scale height) so that it takes
a peanut-shaped or boxy shape?  In absence of clear-cut dynamical
data on the motions of stars in the bulge, the photometric data alone
cannot choose among these options.  Even so, the nearly pure
exponential nature of the bulge of NGC~4565 hints more of a bar than
it does of a typical R$^{1/4}$--law bulge.

\begin{figure}
\figurenum{11}
\centerline{\psfig{file=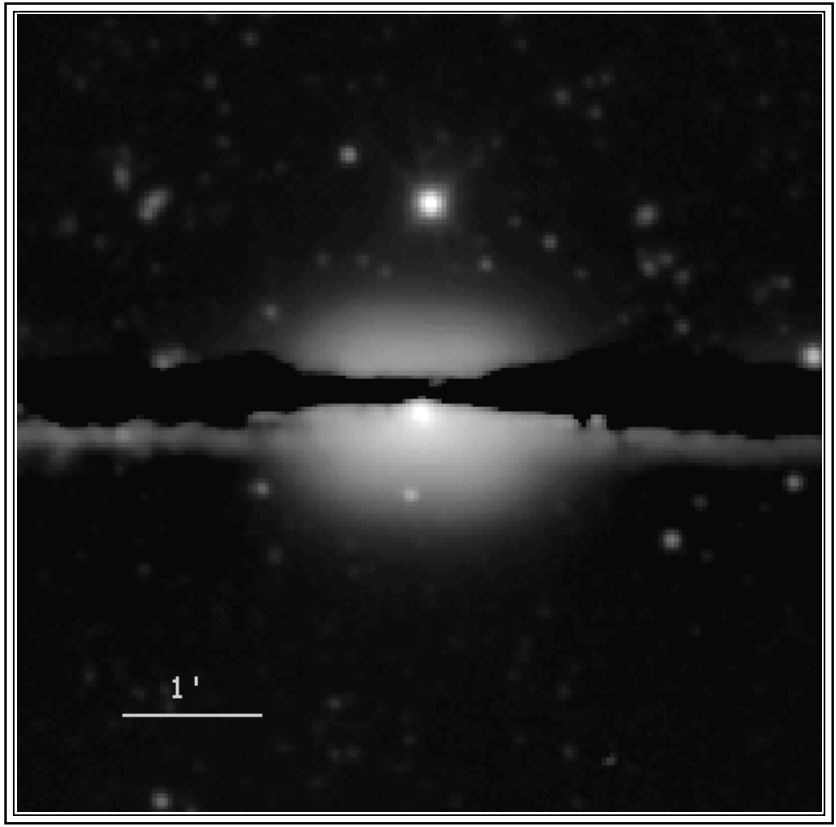,width=12.0cm}}
\centerline{\psfig{file=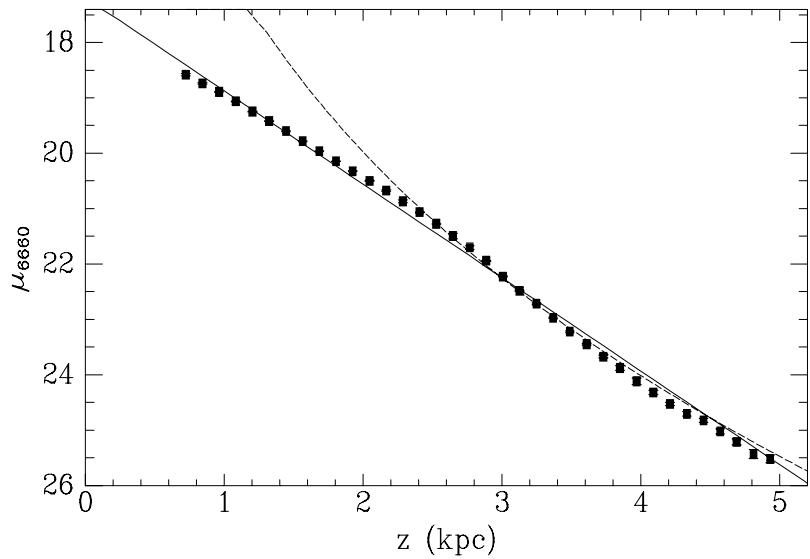,width=12.0cm}}
\vspace{-1cm}
\caption[]{Top: The image of the bulge of NGC~4565 produced
by subtracting the two-disk-halo model from the final image.  The
faint stellar images around the bulge are likely globular clusters in
NGC~4565. Bottom: The minor axis profile (solid boxes)
of the bulge in NGC 4565 so-obtained.  The solid line represents
exponential model with the scale height 0.65kpc.  The dashed line is
the R$^{1/4}$ law that best fits the outer part of the bulge.  The
bulge of NGC 4565 is fit better by an exponential model than an $\rm
R^{1/4}$--law model.
\label{fig11}}
\end{figure}

\subsection{Face-On Warp?}

The striking image presented of the bulge in Figure~11 highlights an
issue that also can be seen in sky-subtracted image of the whole
galaxy in Figure~4: there seems to be far fewer faint objects in the
lower SW side of the galaxy disk than in the upper NE side of the galaxy
disk.  As seen in Figure~4, this asymmetry in number of giant objects
seems to extend to 2-3 arcmin in the SW direction.  Unfortunately, we
cannot appeal to HST observations of this galaxy, as such imaging
(\cite{KP99}) avoided the NE side owing to the bright star there.

The only reasonable interpretation we can give to the existence
of an asymmetry in the faint objects around NGC~4565 is that there
is obscuring matter between NGC~4565 and the background galaxies.
If this proves to be true (and we plan to investigate this issue
further in another paper), then it could be evidence for a face-on
warp in this galaxy.  Such a warp would have to be part of the disk 
facing away from us, as the nucleus of this galaxy is obviously not 
obscured.

\section{SUMMARY}

We have obtained deep intermediate-band ($\rm 6660\AA$) surface
photometry of the nearby, bright, edge-on galaxy NGC 4565. The
combination of having a nearly 1 square degree field of view, the
ability to use the dome to obtain high S/N flat fields, accurate
subtraction of the PSF wings of stars, and an accurate method for
two-dimensional sky subtraction combine to yield a limiting magnitude
of 28.77 mag $\rm arcsec^{-2}$ (at which the observational error
reaches 0.75 mag $\rm arcsec^{-2}$; or 0.25 mag at a surface
brightness of 27.5 mag arcsec$^{-2}$).  The sky background in our image
is 20.72 mag arcsec$^{-1}$, about 0.5 mag arcsec$^{-2}$ brighter than
the sky found in our NGC~5907 observations in the same filter
(cf. Zheng et al. 1999).  The total magnitude of NGC~4565 in 6660$\rm
\AA$-band is 8.99$\pm$0.02 mag (equivalent to R = 9.10) to a 
surface brightness of 28 mag arcsec$^{-2}$.  

The luminosity distribution of the galaxy is presented in a series of
cuts both parallel to its minor axis ($z$-profiles) and parallel to
its major axis ($R$-profiles).  Excluding the dust lane, the
$z$-profiles are symmetric in all four quadrants out to a radius of 6
arcmin.  The $z$-profiles extend to nearly 30 kpc (at 29 mag $\rm
arcsec^{-2}$), farther than given by previous studies of this galaxy.

We construct a two-dimensional model of the outer parts of this galaxy
comprised of three components (thin-disk + thick-disk + halo).
Altogether the model includes 12 parameters, including a cut-off
radius for the disk.  The values of these parameters are determined by
the $\chi^2$ values from a series of 3,700 initial starting values
which eventually converge to distinct values.  The parameters
so-derived for the thin and thick disks generally agree with those
derived by previous studies.  As our halo observations go much deeper
than those of previous studies, we obtain a more reliable measurement
of the power--law behavior of the halo, with its power-law index
$\gamma$ determined to lie between 3.2 and 4.0 (best-fit value of
3.88).  Our observations effectively rule out the possibility of an
$r^{-2}$ halo in NGC~4565.  We obtain an axis ratio for the halo of
this galaxy to be equal to 0.44 and core radius of 14.4 kpc, which
suggest a flattened ellipsoid.

We test for two factors which could affect our results: scattered light 
from the galaxy itself and an incorrect value used for the inclination 
of the  galaxy disk.  We find neither significantly affects our
results. However, if our data are accurate enough, the effects 
would be present, at most, at the 0.2--0.3 mag arcsec$^{-2}$ level.
We measure an inclination of NGC~4565 of $i = 87.5^{\circ}$
from the offset of the dust lane in the inner disk from the nucleus
of the galaxy.

We obtain a two-dimensional luminosity distribution for the bulge of
NGC~4565 by subtracting our two-disk+halo model from the original
image. The luminosity profile along the minor axis of the bulge of
NGC~4565 is well fitted by exponential model with scale height similar
to that of the thin disk (consistent with was has been found from
previous studies).

In the coming years, our survey will obtain deep images in all of the
15 BATC passbands (cf. Zhou et al.) of NGC~4565, as well as for a
number of other, nearby galaxies.  With a complete sampling of the
visible and near-infrared spectrum of these galaxies, we will be able
to study the stellar populations in their disks, bulges and halos.

\acknowledgments

We thanks Professors Qiuhe Peng, Dr. Xiangping Wu and Dr. Shude Mao
for valuable discussion.  Also many thanks to Dr. Jianyan Wei.  The
BATC Survey is supported by the Chinese Academy of Sciences (CAS), the
Chinese National Natural Science Foundation (CNNSF) and the Chinese
State Committee of Sciences and Technology (CSCST). We also thank the
Chinese National Pandeng Project for financial support. The project
was also supported in part by the U.S. National Science Foundation
(NSF Grant INT-93-01805), by Arizona State University, the University
of Arizona, and Western Connecticut State University.

\end{document}